\documentclass[aps,prd,reprint,showpacs]{revtex4-1}
\usepackage{amssymb}
\usepackage{amsmath}
\usepackage{graphicx}

\graphicspath{{figures/}}

\begin{document}

%		setting options for natbib
\setcitestyle{numbers,square,comma}
%\biboptions{numbers,square,comma,merge}

\defcitealias{Shafi:2010jr}{Paper~I}
%\citetalias{Shafi:2010jr}

\title{Observable Gravity Waves from Supersymmetric Hybrid Inflation II}

\author{Mansoor Ur Rehman}\email{rehman@udel.edu}
\author{Qaisar Shafi}\email{shafi@bartol.udel.edu}
\author{Joshua R. Wickman}\email{jwickman@udel.edu}
\affiliation{Bartol Research Institute, Department of Physics and Astronomy, 
University of Delaware, Newark, Delaware 19716, USA}

\begin{abstract}

It is shown that a tensor-to-scalar ratio close to $r = 0.03$, which can be observed by Planck, is realized in supersymmetric hybrid inflation models with TeV-scale soft supersymmetry breaking terms. This extends our previous analysis, which also found $r \lesssim 0.03$ but employed intermediate scale soft terms. Other cosmological observables such as the scalar spectral index are in good agreement with the WMAP data.

\end{abstract}


\maketitle

%%%%%		Introduction

Models of cosmic inflation constructed within the pervasive framework of supersymmetry (SUSY) are attractive in the current era of TeV-scale experiments, such as the Large Hadron Collider (LHC), which are poised to search for and probe SUSY dynamics.  In the early universe, the possible existence and nature of a grand unified theory (GUT) is also expected to have played an important role.  SUSY hybrid inflation provides an elegant framework in which both inflation and GUT symmetry breaking can be simultaneously achieved within the context of SUSY, and can be brought into agreement with current cosmological evidence via the inclusion of various well-motivated corrections.  Methods of obtaining a red-tilted power spectrum (i.e. with scalar spectral index $n_s < 1$) are, by now, well studied, and include (for example) the use of a non-minimal K\"ahler potential~\citep{BasteroGil:2006cm,urRehman:2006hu} or soft SUSY-breaking terms~\citep{Rehman:2009nq,Rehman:2009yj}.  On the other hand, such models typically predict tensor fluctuation amplitudes far below the expected sensitivity of the current Planck satellite experiment.  A precise measurement of this amplitude promises to yield valuable information, such as the abundance of primordial gravitational waves and the energy scale of inflation.

In a recent article, we have revealed a region of the SUSY hybrid parameter space which can lead to significantly larger tensor amplitudes with a tensor-to-scalar ratio as high as $r \simeq 0.03$~\cite{*[{}] [{, and references therein.  Submitted to Physics Letters B.}] Shafi:2010jr}.  (For convenience, we will refer to this article as \citetalias{Shafi:2010jr}.)  This number tantalizingly (and perhaps fortuitously) grazes the edge of values that may be measurable by Planck, and can readily be obtained for the WMAP 7-yr central value $n_s \approx 0.968$ for the spectral index~\citep{Komatsu:2010fb}.  To achieve these larger values of $r$ we have determined that, within the set of assumptions put forth in that treatment, it is necessary to employ supergravity (SUGRA) with sizable (yet perturbative) values of the couplings in a non-minimal K\"ahler potential, as well as an intermediate-scale soft SUSY-breaking mass for the inflaton.  In SUSY model building, it is easier and perhaps better motivated to generate superparticle masses closer to the TeV scale, a characteristic mass scale which is typically built into SUSY models in connection with solving the gauge hierarchy problem.  In this sense, it would be advantageous if large tensor amplitudes can be generated in the presence of TeV-scale soft SUSY-breaking masses.  As we will show, this scenario becomes possible by relaxing only one of our previous assumptions, namely that the coefficient $\kappa_S$ of the term quartic in the inflaton field in the K\"ahler potential is positive.

%%%%%		Background

SUSY hybrid inflation is described by the superpotential~\citep{Dvali:1994ms,Copeland:1994vg}
\begin{equation} 
W = \kappa S(\Phi \overline{\Phi} - M^{2}) \, ,
\label{superpot}
\end{equation}
where $S$ is a gauge singlet superfield containing the inflaton scalar, $\Phi, \overline{\Phi}$ transform nontrivially under some gauge group $G$, and $M$ is a mass parameter corresponding to the scale at which $G$ is broken.  With this field content, Eq.~(\ref{superpot}) is the unique renormalizable superpotential consistent with both $G$ and an additional U(1) `$R$-symmetry,' which is commonly employed to suppress proton decay.  In this article, we will choose $G \equiv \text{U(1)}$, which may be identified with a $B-L$ gauge symmetry.  Making a different choice of gauge group does not substantially affect the empirical results, but may be advantageous in order to address issues with cosmic strings in the U(1) case.  The group SU(5)$\times$U(1), which describes the flipped SU(5) model, constitutes a particularly favorable choice; see \citetalias{Shafi:2010jr} for more details.

The K\"ahler potential may be expanded as
\begin{multline}
K = |S|^2 + |\Phi|^2 + |\overline{\Phi}|^2 + \frac{\kappa_S}{4}\frac{|S|^4}{m_P^2} + \frac{\kappa_\Phi}{4}\frac{|\Phi|^4}{m_P^2} +\frac{\kappa_{\overline{\Phi}}}{4}\frac{|\overline{\Phi}|^4}{m_P^2} \\
 + \kappa_{S \Phi}\frac{|S|^2|\Phi|^2}{m_P^2} + \kappa_{S \overline{\Phi}}\frac{|S|^2|\overline{\Phi}|^2}{m_P^2} + \kappa_{\Phi \overline{\Phi}}\frac{|\Phi|^2|\overline{\Phi}|^2}{m_P^2} \\
 + \frac{\kappa_{SS}}{6}\frac{|S|^6}{m_P^4} + \cdots ,
\label{kahler}
\end{multline}
where $m_P \approx 2.4 \times 10^{18}$~GeV is the reduced Planck mass.  Along the $D$-flat direction, the gauge conjugate fields are related by $|\Phi|=|\overline{\Phi}|$, and during inflation the scalar components of the fields settle into their instantaneous vacuum expectation value $s > M$, $\phi = \overline{\phi} = 0$.  The $F$-term supergravity (SUGRA) scalar potential during inflation then becomes
\begin{equation}
V_F = \kappa^2 M^4 \left( 1 - \kappa_S \frac{|s|^2}{m_P^2} + \frac{1}{2} \gamma_S \frac{|s|^4}{m_P^4} + \cdots \right) ,
\nonumber
\label{Vsugra}
\end{equation}
where 
\begin{equation}
\gamma_S \equiv 1 - \frac{7}{2}\kappa_S + 2\kappa_S^2 - 3\kappa_{SS}.
\label{gammaS}
\end{equation}
Since SUSY is broken along the inflationary trajectory, radiative and soft SUSY-breaking corrections can also modify the inflationary potential.  Including these and SUGRA corrections up to order $|s|^4$, the inflationary scalar potential becomes
\begin{multline}
V \simeq \kappa^{2}M^{4} \left( 1 - \kappa_S \left( \frac{M}{m_P} \right)^2 x^2 + \gamma_S\left( \frac{M}{m_{P}}\right)^{4}\frac{x^{4}}{2} \right. \\
 \left. + \frac{\kappa ^{2}\mathcal{N}}{8\pi ^{2}}F(x) + a\left(\frac{m_{3/2}\,x}{\kappa\,M}\right) + \left( \frac{M_S\,x}{\kappa\,M}\right)^2\right) ,
\label{scalarpot}
\end{multline}
where we have used the convenient parametrization $x \equiv |s|/M$ of the inflaton field.  The function~\citep{Dvali:1994ms}
%\begin{equation}
\begin{multline}
F(x)=\frac{1}{4}\left( \left( x^{4}+1\right) \ln \frac{\left( x^{4}-1\right)}{x^{4}}+2x^{2}\ln \frac{x^{2}+1}{x^{2}-1} \right. \\
\left. + 2\ln \frac{\kappa ^{2}M^{2}x^{2}}{Q^{2}}-3\right)
\end{multline}
%\end{equation}
embodies the radiative corrections, and $a$ and $M_S$ are the effective coefficients of the soft SUSY-breaking linear and mass terms for the inflaton, respectively.  In the simplest models, $\mathcal{N}$ is the dimensionality of $\Phi, \overline{\Phi}$ under $G$ ($\mathcal{N}=1$ in the U(1) case, and $\mathcal{N}=10$ for flipped SU(5)), and $Q$ is the renormalization scale.  In a gravity-mediated SUSY breaking scheme, the gravitino mass is taken to be $m_{3/2} \sim 1$~TeV, and as stated previously, we are interested in the case where the soft SUSY-breaking mass takes on comparable values, $M_S \sim m_{3/2}$ (with $M_S^2 > 0$).

In contrast to the previous treatment, we take $\kappa_S < 0$.  As noted in \citetalias{Shafi:2010jr}, this change (while maintaining $\kappa_{SS} > 0$) may result in the same large-$r$ behavior without requiring intermediate soft SUSY-breaking mass scales $M_S \gg $~TeV.  Eq.~(\ref{gammaS}) then tells us that the coefficient $\gamma_S$ of the quartic term in the potential only becomes negative if $\kappa_{SS}$ is at least $\sim 1/3$.  We recall that $\gamma_S < 0$ is typical of points having the largest $r$-values.

Without loss of generality, we choose $\kappa > 0$.  In order to ensure that SUGRA remains under control, we limit the field amplitude to $|s| \leq m_P$.  We utilize the same calculational approach, assumptions, and parameter ranges as put forth in \citetalias{Shafi:2010jr}.  It may be pertinent to draw the reader's attention to one of these assumptions, namely that the potential is monotonic along the inflationary trajectory.  While nontrivial in nature, this assumption was instituted to ensure that no metastable vacua appear that may ruin the inflationary evolution.

%%%%%		Analytic approximation

With $M_S \sim 1$~TeV, the soft SUSY-breaking mass-squared term is strongly suppressed relative to the SUGRA quadratic term for the entire allowed range of $|\kappa_S|$.  If $a$ is taken to be order unity or smaller, the linear term is also small for $y \equiv |s|/m_P \sim 1$ (i.e. for the largest values of $r$).  Then, the asymptotic form of $r$ obtained in \citetalias{Shafi:2010jr} remains intact, and may be written as
\begin{eqnarray}
r &\simeq& 16 y_0^2 \left[ |\kappa_S| + \gamma_S y_0^2 \right]^2 , \label{rapprox1} \\
  &\simeq& 16 y_0^2 \left[ \frac{2}{3}|\kappa_S| + \frac{1}{3}\eta \right]^2 . \label{rapprox2}
\end{eqnarray}
(Here, we have suppressed the radiative correction term, which is found to be unimportant at large $y$ under most circumstances.)  Using the usual slow-roll formulae $n_s \simeq 1 - 6\epsilon + 2\eta$ and $r \simeq 16\epsilon$, we may trade $\eta$ in Eq.~(\ref{rapprox2}) for $n_s$, yielding
\begin{multline}
r_\pm \simeq \frac{8}{3 y_0^2} \left[ 3 - 4|\kappa_S|y_0^2 + (1-n_s)y_0^2 \right. \\
 \left. \pm \sqrt{3 \left( 3 - 8|\kappa_S|y_0^2 + 2(1-n_s)y_0^2 \right)} \right] .
\label{rpm}
\end{multline}
In addition to the solution $r_-$ comprising an excellent approximation to the calculated $r$-values, Eq.~(\ref{rpm}) elucidates the relation between $r$ and $|\kappa_S|$.  Fixing $y_0 = 1$ and $n_s = 0.968$, the highest values $r \approx 0.03$ observed in our numerical results can be obtained for $|\kappa_S| \approx 0.07$.

In the analysis above, we have argued that the radiative correction term may be neglected for the largest $r$-values.  However, while this term does not directly play a substantial role in the expressions for quantities such as $r$ and $n_s$, it can become important at the end of inflation, indirectly via the number of e-foldings $N_0$.  In terms of $y$, the potential may be approximated as
\begin{equation}
V(y) \approx V_0 \left( 1 + |\kappa_S|\,y^2 + \gamma_S  \frac{y^4}{2} + \frac{\kappa^2\,\mathcal{N}}{8\,\pi^2}\ln \frac{\kappa \, m_P \, y}{Q} \right),
\end{equation}
where $V_0 \equiv \kappa^2 M^4$, and we have written the radiative correction term in an asymptotic form for large $x$ (or, more appropriately in this case, large $\kappa$).  The number of e-foldings is then given by the integral
\begin{eqnarray}
N_0 &\simeq& 2 \int_{\frac{M}{m_P}}^{1} \left( \frac{V}{\partial_y V} \right) dy \nonumber \\
 &\approx& \int_{\frac{M}{m_P}}^{1}  \frac{dy}{y\left(|\kappa_S| +\gamma_S y^2
+ \frac{\kappa^2\,\mathcal{N}}{(4\,\pi)^2} \frac{1}{y^2}\right)},
\label{N0}
\end{eqnarray}
where we have taken inflation to end via waterfall (corresponding to $x=1$).  For most of the inflationary trajectory, $y$ is large enough so that radiative corrections are suppressed relative to the other terms unless $\kappa$ is quite large.  In this region, an increase in $\kappa$ amounts to a decrease in $M/m_P$, i.e. a widening of the integration interval.  This in turn tends to increase $N_0$, and must be accompanied by a decrease in the integrand in order for $N_0$ to remain within the usual window 50--60.  Even for large $y$, the $|\kappa_S|$ term must dominate over the $\gamma_S$ term in order for our assumption $V'>0$ to remain intact.  Then, an increase in $\kappa$ should be mitigated by an increase in $|\kappa_S|$ in the region where radiative corrections are not important.

\begin{figure}
\includegraphics[width=\columnwidth]{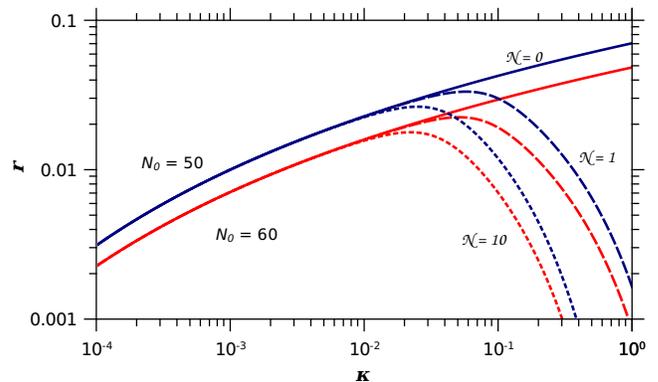}
\caption{The behavior of $r$ with respect to $\kappa$ for $n_s$ fixed at the WMAP7 central value and $y_0$ fixed at 1 (Planck scale).  The blue (red) curves correspond to $N_0=50$~(60), and different choices of gauge group corresponding to $\mathcal{N} = 1,\, 10$ are shown using dashed and dotted curves, respectively.  For comparison, the case with no radiative corrections ($\mathcal{N} = 0$, solid curves) is also shown.  Note that $\mathcal{N} = 10$ describes the flipped SU(5) model.}
\label{r-k_approx}
\end{figure}

At sufficiently large $\kappa$, the radiative correction term becomes large enough to have an influence on the integrand.  In this regime, an increase in $\kappa$ leads directly to a decrease in the integrand even for constant $|\kappa_S|$.  Then the requirement that $|\kappa_S|$ increase is relaxed, and for very large $\kappa$ we will even begin to see a decrease in $|\kappa_S|$ in order for $N_0$ not to become too small.  Naturally, this argument implies that there exists some critical range of $\kappa$ where $|\kappa_S|$ ceases to increase and begins to decrease; in other words, due to radiative corrections, there is an upper bound on the value of $|\kappa_S|$.  This bounding value will vary with parameters such as $n_s$, $y_0$, and $N_0$.  Of particular interest is the impact of this prediction on the behavior of $r$.  If we consider the approximate form $r_-$ in Eq.~(\ref{rpm}) for fixed $n_s$ and $y_0=1$, we see that critical behavior of $|\kappa_S|$ leads to the same conclusion for $r$, i.e. $d|\kappa_S|/d\kappa=0$ implies $dr/d\kappa=0$.  This behavior is displayed in Fig.~\ref{r-k_approx}.  By continuing the analysis above, the maximal value of $r$ can be written as
\begin{equation}
r_m^{5/4}= \left(\frac{m_P}{M_G}\right)^2
\left(\frac{(4\,|\kappa_{S_m}|)^2}{\sqrt{27\,\mathcal{N}/(4\,\pi)^2}}\right) 
e^{-3\,|\kappa_{S_m}|\,N_0},
\label{rm}
\end{equation}
where a subscript $m$ denotes a value at the maximum of $r$ (or $|\kappa_S|$) with respect to $\kappa$, and $M_G = 3.3 \times 10^{16}$~GeV represents the GUT scale.  Numerically solving Eqs.~(\ref{rpm}) and (\ref{rm}) simultaneously for fixed values of $n_s,\, y_0,\, \mathcal{N},\, N_0$ yields a prediction for the bounding values of $r$ and $|\kappa_S|$.  Taking $n_s=0.968,\, y_0=1,\, \mathcal{N}=1,\, N_0=50$ gives $|\kappa_{S_m}| \approx 0.08$, $r_m \approx 0.04$.  We will see that this lies very close to the values seen in our numerical results.

%%%

%%%

%%%%%		Numerical results and discussion

\begin{figure}
\includegraphics[width=\columnwidth]{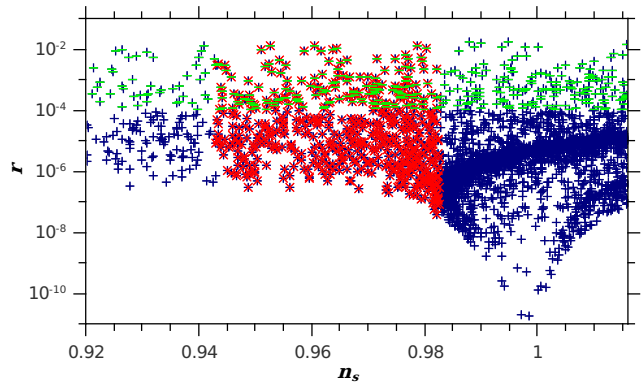}
\caption{The $r$~vs.~$n_s$ plane for $n_s$ within a 4$\sigma$ range of the central value.  The full set of numerical results (blue crosses) is overlaid by two subsets: 1) the region contained within the WMAP7 1$\sigma$ bounding curves in this plane (red X's), and 2) points having $r \gtrsim 10^{-4}$ (green lines).  The largest values of $r$, saturating the analytical bound and potentially observable by the Planck satellite, can be achieved for any $n_s$ of interest.}
\label{r-ns}
\end{figure}

\begin{figure}
\includegraphics[width=\columnwidth]{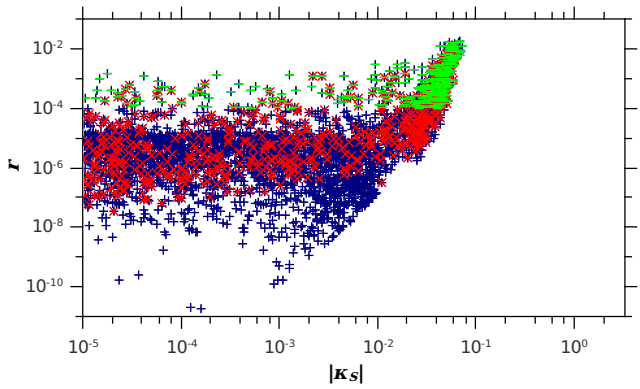}
\caption{The tensor-to-scalar ratio $r$ as a function of $|\kappa_S|$.  The color and symbol coding is the same as in Fig.~\ref{r-ns}.  The behavior seen here is similar to that of the effective quadratic coefficient in the $M_S \gg 1$~TeV case (see \citetalias{Shafi:2010jr}).}
\label{r-kS}
\end{figure}

The results of our full numerical calculations are very similar to those presented in \citetalias{Shafi:2010jr}.  In particular, we again find that $r \lesssim 0.03$ in these models of SUSY hybrid inflation~\footnote{It is interesting to compare this result with that of a non-SUSY hybrid inflation model, where radiative corrections are important in obtaining $r \lesssim 0.004$ (for $n_s \approx 0.968$) with sub-Planckian field values~\citep{Rehman:2009wv}.}, and that this value can readily be achieved within the 1$\sigma$ region of the $(n_s, r)$ plane as delineated by the WMAP7 analysis~\citep{Komatsu:2010fb}.  These results are shown in Fig.~\ref{r-ns}.  Presented in Fig.~\ref{r-kS} is the behavior of $r$ with respect to $|\kappa_S|$; qualitatively, the relation between these parameters looks very similar to the behavior of $r$ with respect to $M_S$ in \citetalias{Shafi:2010jr}.  This can be understood by noting that $|\kappa_S|$ plays essentially the same role in the $M_S \sim $~TeV case as was played by $M_S$ in the $M_S \gg $~TeV case, namely that of the (positive) effective quadratic coefficient in the potential.

%%%

%%%

%%%

%%%

\begin{figure}
\includegraphics[width=\columnwidth]{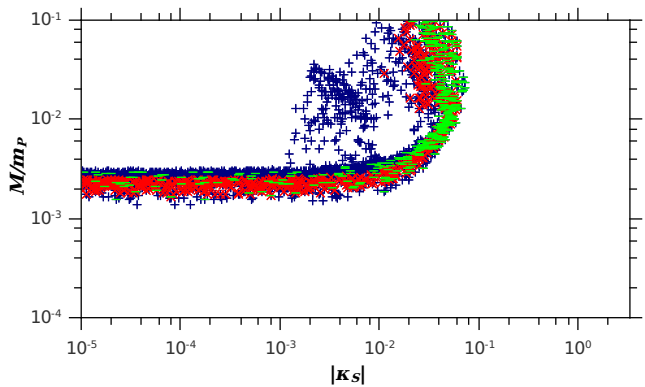}
\caption{The gauge symmetry breaking scale $M$ as a function of $|\kappa_S|$.  The color and symbol coding is the same as in Fig.~\ref{r-ns}.  For small $|\kappa_S|$, the model reduces to that having only radiative corrections, which predicts $M$ around the GUT scale.  If SUGRA corrections are important, larger values of $M$ become possible.}
\label{M-kS}
\end{figure}

Another interesting feature emerging from plots of our results is a distinct limit on the value of the symmetry breaking scale, namely $M \gtrsim 10^{16}$~GeV.  This bound is saturated in the limit $\kappa_S, \gamma_S \rightarrow 0$, in other words, in the limit of no SUGRA corrections.  Since the soft SUSY-breaking terms are also highly suppressed, the potential is essentially comprised of only the vacuum and radiative correction terms, which coincides with the minimal model of SUSY hybrid inflation described in Ref.~\citep{Dvali:1994ms}.  This model predicts that the temperature anisotropy of the Cosmic Microwave Background (CMB) turns out to be $\delta T/T \sim (M/m_P)^2$, which in turn demands that $M$ be of order the GUT symmetry breaking scale.  It turns out that when SUGRA corrections are introduced, the slope of the inflationary trajectory can be large enough to produce enough e-foldings without a large contribution from radiative corrections, and $\kappa$ may be relaxed to smaller values.  Since the vacuum scale $V_0^{1/4}$ does not change much, this leads to an increase of $M$ in the case of substantial SUGRA corrections.  The behavior of $M$ as a function of $|\kappa_S|$ is shown in Fig.~\ref{M-kS}.  It is worth noting that this conclusion also applies to the model in \citetalias{Shafi:2010jr}, yet modified by the additional variation of $M_S$.

%%%

%%%

In addition to the new results described so far, the predictions of \citetalias{Shafi:2010jr} remain largely intact.  For example, the running of the spectral index $|dn_s/d\ln k|$ can again be about $\sim 0.01$ at the largest $r$-values, and much smaller for lower $r$, in agreement with the experimental observations.  Also, the predictions of the model again change only modestly if the flipped SU(5) group is chosen for $G$, and so we have chosen not to display those results here.  Accordingly, the possible issues addressed in \citetalias{Shafi:2010jr} have not acquired any fortuitous solutions in the present case, and we refer the reader there for a detailed discussion.

%%%%%		Summary

In summary, we have extended the analysis of \citetalias{Shafi:2010jr} to treat the case with $\kappa_S < 0$ in the K\"ahler potential, and shown that using soft SUSY-breaking masses of order $\sim 1$~TeV, the same predictions can be realized.  In particular, this means that models of SUSY hybrid inflation can support large tensor modes without invoking intermediate mass scales as previously thought, eliminating the need for a specialized mechanism to generate such masses and thus enhancing the attractiveness of the model.  If $\kappa_S < 0$ is chosen, only the non-minimal K\"ahler potential is needed to access the region of parameter space leading to large tensor modes.  We have also discussed additional behavior in the model, including an upper bound placed on $r$ due in part to radiative corrections.  The tensor-to-scalar ratio can take on values up to $\sim 0.03$, fortuitously close to the values accessible by the current Planck satellite.  This can be achieved for the spectral index at its central value $n_s \approx 0.968$; other cosmological observables are also in agreement with the latest WMAP7 data.  The flipped SU(5) model provides a compelling example of the hybrid inflation scenario we have presented.

\begin{acknowledgments}

This work is supported in part by the DOE under grant No.~DE-FG02-91ER40626, by the University of Delaware Dissertation Fellows award (M.R.), and by NASA and the Delaware Space Grant Consortium under grant No.\ NNG05GO92H (J.W.).

\end{acknowledgments}

\bibliographystyle{apsrev4-1}%{h-physrev}
\bibliography{tevscanref}

\end{document}